\documentclass[epj]{svjour}

\usepackage{a4wide}
\usepackage{graphicx}
\usepackage{amsfonts}
\usepackage{amsmath}
\usepackage{amssymb}
\usepackage{booktabs}
\usepackage[colorlinks]{hyperref}
\hypersetup{citecolor=blue,linkcolor= blue}
\begin{document}
\title{Testing power-law cross-correlations: Rescaled covariance test}
\author{Ladislav Kristoufek\inst{1,2}}
\institute{Institute of Information Theory and Automation, Academy of Sciences of the Czech Republic, Pod Vodarenskou Vezi 4, 182 08, Prague, Czech Republic, EU
 \and Institute of Economic Studies, Faculty of Social Sciences, Charles University in Prague, Opletalova 26, 110 00, Prague, Czech Republic, EU
}
\date{Received: date / Revised version: date}
%
\abstract{
We introduce a new test for detection of power-law cross-correlations among a pair of time series -- the rescaled covariance test. The test is based on a power-law divergence of the covariance of the partial sums of the long-range cross-correlated processes. Utilizing a heteroskedasticity and auto-correlation robust estimator of the long-term covariance, we develop a test with desirable statistical properties which is well able to distinguish between short- and long-range cross-correlations. Such test should be used as a starting point in the analysis of long-range cross-correlations prior to an estimation of bivariate long-term memory parameters. As an application, we show that the relationship between volatility and traded volume, and volatility and returns in the financial markets can be labeled as the power-law cross-correlated one.
\PACS{
      {05.10.-a}{Computational methods in statistical physics and nonlinear dynamics}   \and
      {05.45.-a}{Nonlinear dynamics and chaos}   \and
      {89.65.Gh}{Economics; econophysics, financial markets, business and management}
     } 
} 
\authorrunning{Kristoufek}
\maketitle

\section{Introduction}
Analysis of the power-law auto-correlations and long-term memory has a long tradition in the econophysics field. Starting from the early studies in 1990s \cite{Mantegna1999,Liu1999,Gopikrishnan2000,Plerou2000}, the main focus has been put on financial time series, specifically scaling of auto-correlations of returns and volatility measures. The long-range dependent processes are characterized by the long-term memory parameter $H$ -- Hurst exponent -- which ranges between 0 and 1 for stationary processes. The breaking point of 0.5 is characteristic for uncorrelated and short-term memory processes (with exponentially decaying auto-correlations). Processes with $H>0.5$ are labeled as persistent and they resemble locally trending series, and processes with $H<0.5$ are anti-persistent with frequently switching direction of increments. The dynamics of the long-term dependent series with $H\ne 0.5$ is pronounced in the scaling of the auto-correlation function $\rho(k)$ with lag $k$ which follows an asymptotic power-law decay, $\rho(k)\propto k^{2H-2}$ for $k \rightarrow \pm \infty$, and in the divergence of the spectrum $f(\lambda)$ with frequency $\lambda$ so that $f(\lambda) \propto \lambda^{1-2H}$ for $\lambda \rightarrow 0+$ \cite{Beran1994}.

Availability of huge sets of financial data has increased the number of empirical studies and the topic of the power-law scaling of auto-correlation functions remains a popular topic \cite{DiMatteo2007,Power2010,Alvarez-Ramirez2009,Kristoufek2010a,Kristoufek2010b,Fleming2011,Chakraborti2011,Barunik2012}. Apart from the empirical works, there have been numerous papers on statistical properties of various estimators of the long-term memory discussing their performance under various memory and distributional properties \cite{Taqqu1995,Taqqu1996,Taqqu1998,Weron2002,Kantelhardt2002,couillarddavidson2005,grech2005,Barunik2010,Kristoufek2010,Kristoufek2012}. These studies show that practically all estimators are biased by some of these properties and spurious long-term memory can be quite easily reported. Several tests for presence of long-term memory have been proposed as an initial step in the long-term memory analysis. The original rescaled range has been proposed by Hurst \cite{Hurst1951} and later adjusted by Mandebrot \& Wallis \cite{Mandelbrot1968a}. Lo \cite{Lo1991} proposes a modified version of the rescaled range statistic which controls for the short-term memory bias.
Giraitis \textit{et al.} \cite{Giraitis2003} introduce the rescaled variance statistic and show that it supersedes the modified rescaled range analysis and KPSS statistic \cite{Kwiatkowski1992} for various settings of short-term and long-term memory processes.

With the outburst of the Global Financial Crisis in 2007/2008, the study of correlations and cross-correlations between various assets has attracted an increasing interest. In econophysics, growing number of papers has focused on the power-law behavior of the cross-correlation function \cite{Podobnik2009,Podobnik2009a,SiqueiraJr.2010,He2011,He2011b,Ma2013,Wang2013,Wang2013a}. To this point, several estimators of the bivariate Hurst exponent $H_{xy}$ have been introduced -- detrended cross-correlation analysis (DCCA) \cite{Podobnik2008,Zhou2008,Gu2010,Jiang2011}, multifractal height cross-correlation analysis (MF-HXA) \cite{Kristoufek2011}, detrended moving-average cross-correlation analysis (DMCA) \cite{He2011a}, multifractal statistical moments cross-correlation analysis (MFSMXA) \cite{Wang2012} and average periodogram method (APE) \cite{Sela2012}. Compared to the univariate case, there has been practically no attention given to an actual testing for presence of the power-law cross-correlations between two series. Up to our best knowledge, there has been only one test proposed by Podobnik \textit{et al.} \cite{Podobnik2011} utilizing the DCCA-based correlation of Zebende \cite{Zebende2011}. 

We propose a new test based on the divergence of covariance of the partial sums of the power-law cross-correlated processes which is robust to short-term memory effects -- the rescaled covariance test. The paper is structured as follows. In Section 2, basic definitions and concepts of the long-range cross-correlated processes are introduced together with propositions needed for the construction of the rescaled covariance test in Section 3. Finite sample properties of the test are described in Section 4. In Section 5, the test is applied on a set of financial time series. Section 6 concludes.

\section{Methodology}

The power-law (or long-term/long-range) cross-correlated processes can be defined in multiple ways -- to name the most important ones, via scaling of the cross-correlation function or a slowly at infinity varying function, through a non-summability of the cross-correlation function, and a divergent at origin cross-power spectrum. For our purposes, it is sufficient to define the long-range cross-correlated processes via the power-law decay of the cross-correlation function $\rho_{xy}(k)$ with time lag $k \in \mathbb{Z}$ defined as
\begin{equation}
\rho_{xy}(k)=\frac{\langle (x_t-\langle x_t \rangle)(y_{t-k}-\langle y_t \rangle) \rangle}{\sqrt{\langle x^2_t-\langle x_t \rangle^2\rangle \langle y^2_t-\langle y_t \rangle^2\rangle}}.
\end{equation}

The following two definitions illustrate the crucial difference between short-range and long-range cross-correlated processes which stems in a contrast between decay and vanishing of the cross-correlation function.\\

\textbf{Definition: Short-range cross-correlated processes.} Two jointly stationary processes $\{x_t\}$ and $\{y_t\}$ are short-range cross-correlated (SRCC) if for $k>0$ and/or $k<0$, the cross-correlation function behaves as
\begin{equation}
\label{eq:SRCC1}
\rho_{xy}(k)\propto \exp(-k/\delta)
\end{equation}
with a characteristic time decay $0\le \delta<+\infty$. \\

\textbf{Definition: Long-range cross-correlated processes.} Two jointly stationary processes $\{x_t\}$ and $\{y_t\}$ are long-range cross-correlated (LRCC) if for $k\rightarrow +\infty$, the cross-correlation function behaves as
\begin{equation}
\label{eq:LRCC1}
\rho_{xy}(k)\propto k^{-\gamma_{xy}}
\end{equation}
with a long-term memory parameter $0<\gamma_{xy}<1$.\\

The definition of the LRCC process thus needs only a half of the cross-correlation function to follow the power-law and the same is true for the SRCC processes. If the cross-correlation function vanishes exponentially for $k<0$ and decays hyperbolically for $k>0$, it is treated as LRCC as the power-law decay dominates the exponential one. In a more general sense, the cross-correlation function is, in contrast to the auto-correlation function, usually asymmetric. However, we show that the asymmetry does not affect several statistical properties of the LRCC, as well as SRCC, processes. Parallel to the univariate case, we label the LRCC processes as either long-range (long-term) cross-correlated or cross-persistent. Contrary to the univariate case, we can separate the LRCC processes between positively (negatively) long-range (long-term) cross-correlated or positively (negatively) cross-persistent.
For practical purposes, the analysis of the asymptotic behavior of cross-correlation function is rather complicated for finite samples. In the time domain, it turns out that it is usually more convenient to study the behavior of partial sums of the processes.\\

\textbf{Definition: Partial sum.} Let's have a stationary process $\{x_t\}$ with $\langle x_t\rangle=0$ and $\langle x_t^2 \rangle = \sigma_x^2<+\infty$. Partial sum process $\{X_t\}$ is defined as
\label{de:PS}
\begin{equation}
X_t=x_1+x_2+\ldots+x_t=\sum_{i=1}^{t}{x_i}.
\end{equation}

Historically, long-range dependence was analyzed by Hurst \cite{Hurst1951} using the rescaled range analysis \cite{Mandelbrot1968a}, which is based on the assumption that the adjusted rescaled ranges of the partial sums of a zero mean process scale according to a power-law. Other measures of variation have been used alongside the adjusted ranges to study long-term dependence, the most popular being the detrended fluctuation analysis \cite{Peng1993,Peng1994,Kantelhardt2002} and various methods covered by Taqqu \textit{et al.} \cite{Taqqu1995,Taqqu1996,Taqqu1998}. We follow this logic for the long-range cross-correlated processes in the next propositions (proofs are given in the Appendix).\\

\textbf{Proposition: Partial sum covariance scaling.}
Let's have two jointly stationary processes $\{x_t\}$ and $\{y_t\}$ and their respective partial sums $\{X_t\}$ and $\{Y_t\}$. If processes $\{x_t\}$ and $\{y_t\}$ are long-range cross-correlated, the covariance between their partial sums scales as
\begin{equation}
\label{eq:prop_PS}
\text{Cov}(X_n,Y_n)\propto n^{2H_{xy}}
\end{equation}
as $n\rightarrow +\infty$ where $H_{xy}$ is the bivariate Hurst exponent. Moreover, it holds that $H_{xy}=1-\frac{\gamma_{xy}}{2}$.\\

\textbf{Proposition: Diverging limit of covariance of partial sums.}
For two jointly stationary long-range cross-correlated processes, $\{x_t\}$ and $\{y_t\}$ and their respective partial sums $\{X_t\}$ and $\{Y_t\}$, it holds that
\begin{equation}
\label{eq:PSlimit}
\lim_{n\rightarrow +\infty}{\frac{\text{Cov}(X_n,Y_n)}{n}}=+\infty.
\end{equation}

The above divergence is parallel to the divergence of the variance of the partial sums for the long-range dependent processes \cite{Samorodnitsky2006} and can thus be seen as a sign of long-range cross-correlations. However, distinguishing between the short- and long-range cross-correlated processes only makes sense if the diverging limit is not the case for the short-range cross-correlated processes. The following proposition and its proof (in the Appendix) indeed show so.\\

\textbf{Proposition: Converging limit of covariance of partial sums.}
For two jointly stationary short-range cross-correlated processes, $\{x_t\}$ and $\{y_t\}$, and their respective partial sums $\{X_t\}$ and $\{Y_t\}$, the expression
\begin{equation}
\lim_{n\rightarrow +\infty}{\frac{\text{Cov}(X_n,Y_n)}{n}}
\end{equation}
converges.\\

We use these definitions to propose a new test for presence of the power-law cross-correlations between two processes -- the rescaled covariance test.

\section{Rescaled covariance test}

Motivated by the works of Giraitis \textit{et al.} \cite{Giraitis2003} and Lavancier \textit{et al.} \cite{Lavancier2010}, we propose a new test for the presence of long-range cross-correlations between two series. The test, which we call the rescaled covariance test, is based on the scaling of the partial sums covariance and on the diverging limit of the covariance of the partial sums. Before proposing the test itself, we need to define the heteroskedasticity and autocorrelation consistent (HAC) estimator of the cross-covariance $s_{xy,q}$ \cite{Giraitis2003,Lavancier2010}.\\

\textbf{Definition: HAC-estimator of covariance.}
Let processes $\{x_t\}$ and $\{y_t\}$ be jointly stationary with a cross-covariance function $\gamma_{xy}(k)$ for lags $k\in \mathbb{Z}$. The heteroskedasticity and auto-correlation consistent estimator of $\gamma_{xy}(0)$ is defined as
\begin{equation}
\label{eq:HAC_CC}
\widehat{s_{xy,q}}=\sum_{k=-q}^q{\left(1-\frac{|k|}{q+1}\right)\widehat{\gamma_{xy}}(k)},
\end{equation}
where $q$ is a number of lags of the cross-covariance function taken into consideration and the cross-covariances are weighted with the Barlett-kernel weights.\\

The basic idea behind the rescaled covariance test (RCT) is to utilize the divergence of covariances of the partial sums of the long-range cross-correlated processes but also the convergence of the short-range cross-correlated processes and at the same time controlling for different levels of correlations in the case of the short-term memory utilizing $\widehat{s_{xy,q}}$. The rescaled covariance test is then defined as follows:\\

\textbf{Definition: Rescaled covariance test.}
Let processes $\{x_t\}$ and $\{y_t\}$, with $t=1,2,\ldots,T$, be jointly stationary processes with a cross-covariance function $\gamma_{xy}(k)$ for $k \in \mathbb{Z}$ and with respective partial sums $\{X_t\}$ and $\{Y_t\}$. Assuming that $\sum_{k=-\infty}^{+\infty}{\gamma_{xy}(k)}\ne 0$, the rescaled covariance statistic $M_{xy,T}(q)$ is defined as
\begin{equation}
\label{eq:RC}
M_{xy,T}(q)=q^{\widehat{H_x}+\widehat{H_y}-1}\frac{\widehat{\text{Cov}}(X_T,Y_T)}{T\widehat{s_{xy,q}}},
\end{equation} 
where $\widehat{s_{xy,q}}$ is the HAC-estimator of the covariance between $\{x_t\}$ and $\{y_t\}$, $\widehat{\text{Cov}}(X_T,Y_T)$ is the estimated covariance between partial sums $\{X_T\}$ and $\{Y_T\}$, and $\widehat{H_x}$ and $\widehat{H_y}$ are estimated Hurst exponents for separate processes $\{x_t\}$ and $\{y_t\}$, respectively.\\

\begin{figure}[!htbp]
\begin{center}
\begin{tabular}{cc}
\includegraphics[width=70mm]{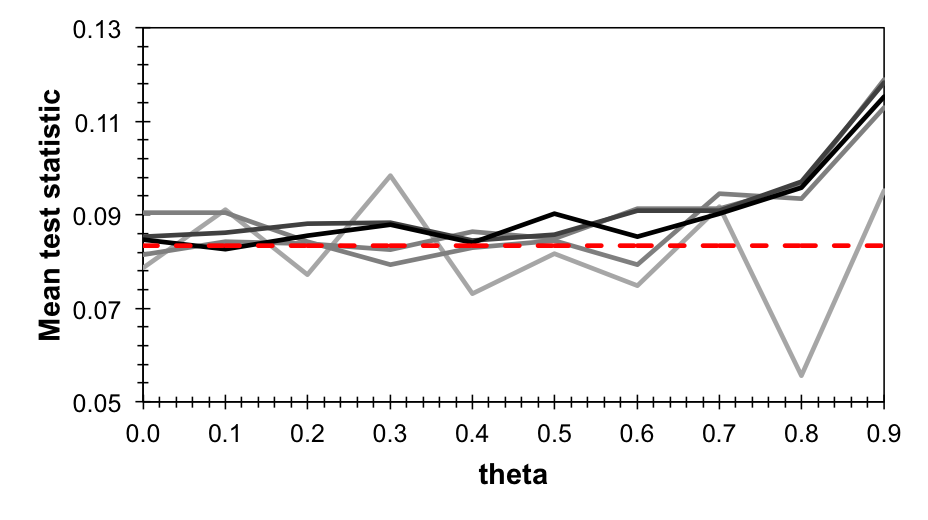}&\includegraphics[width=70mm]{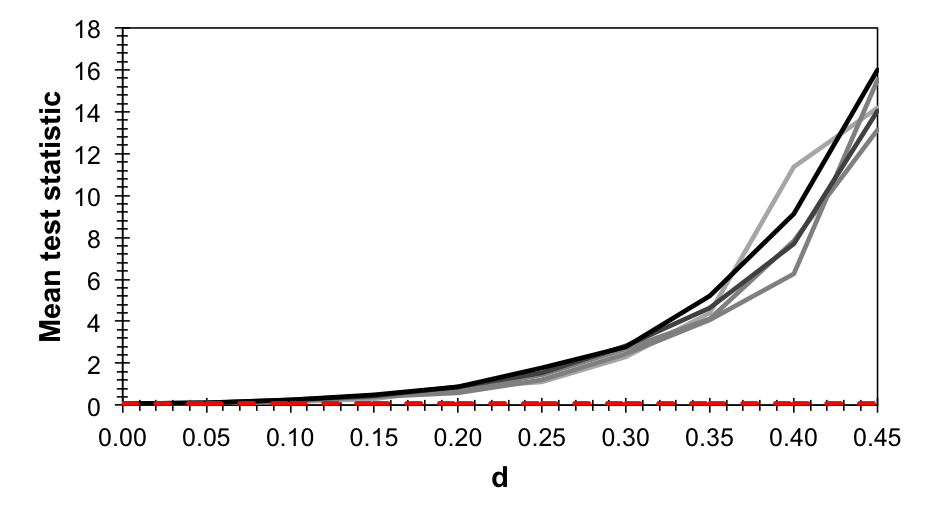}\\
\includegraphics[width=70mm]{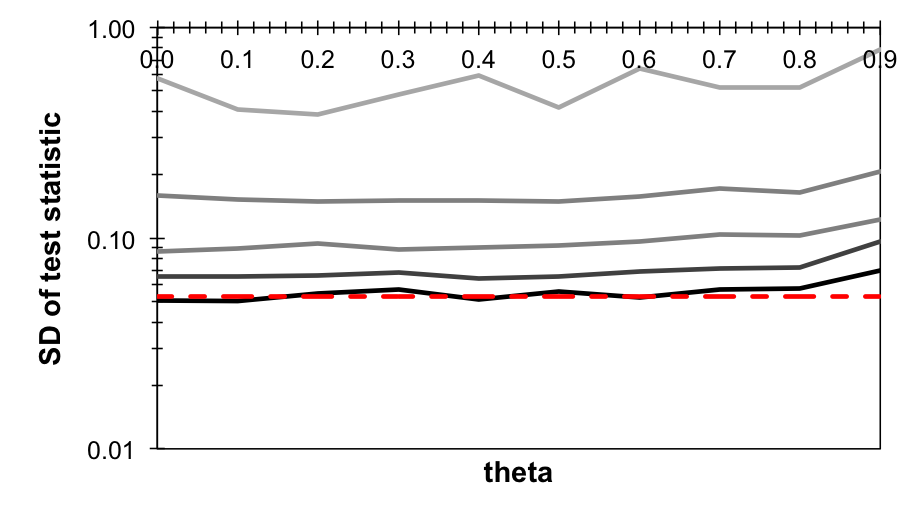}&\includegraphics[width=70mm]{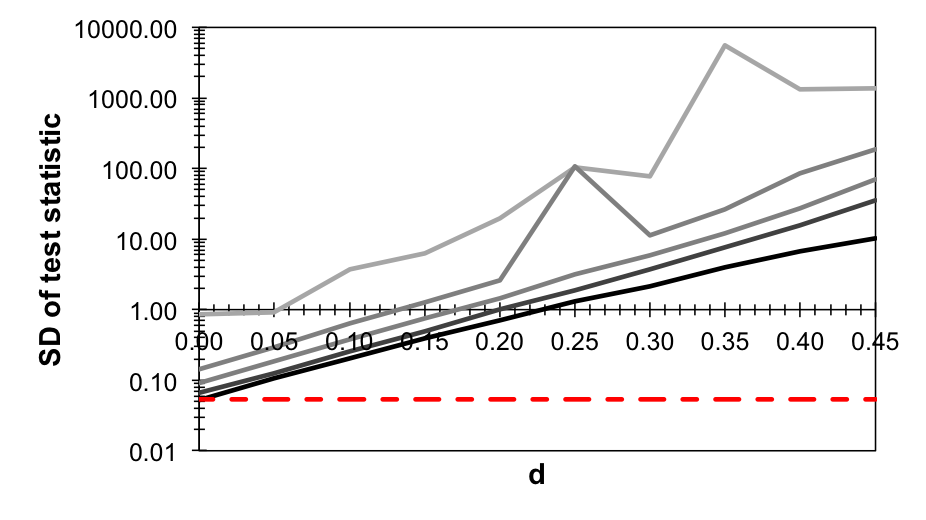}
\end{tabular}
\end{center}
\caption[Mean values and standard deviations of RCT test]{\textbf{Mean values and standard deviations of RCT test.} \footnotesize{Test statistic $M_{xy,5000}(30)$ for differently correlated processes. Correlation between error terms varies between 0.2 and 1 with a step of 0.2 and the darker the line in the chart is, the higher the correlation is. On the left, correlated AR(1) processes with $\theta$ ranging between 0 and 0.9 with a step of 0.1 are shown. On the right, correlated ARFIMA(0,$d$,0) processes with $d$ ranging between 0 and 0.45 with a step of 0.05 are shown. Means are based on 1,000 simulations with a time series length of 5,000 and presented in a semi-log scale for better legibility.}\label{fig:RCT_1}
}
\end{figure}

Similarly to the tests for long-range dependence in the univariate series which are based on the modified variance, such as the rescaled variance \cite{Giraitis2003} and the modified rescaled range analysis \cite{Lo1991}, the choice of parameter $q$ is crucial. If the parameter is too low, the strong short-range cross-correlations can be detected as the long-range cross-correlations and reversely, if the parameter is too high, the true long-range cross-correlations can be filtered out as the short-range ones. This issue is discussed later. Returning to the construction of RCT, the motivation was to construct a test which would have a test statistic that would be (at least partially) independent of the parameters included in the null hypothesis. For the test, we have the null hypothesis of short-range cross-correlated processes and the alternative of cross-persistent processes. Therefore, it is desirable to have a testing statistic independent of the correlation level of the short-range cross-correlated processes as well as the time decay $\delta$. In Fig. \ref{fig:RCT_1}, we present the means and standard deviations of the testing statistics $M_{xy,T}(q)$ for both short- and long-term memory cases with varying parameters. The short-term memory processes are represented by AR(1) processes $\{x_t\}$ and $\{y_t\}$ with correlated error terms and memory parameter $\theta$:
\begin{gather}
x_t=\theta_1x_{t-1}+\varepsilon_t \nonumber \\
y_t=\theta_2x_{t-1}+\nu_t \nonumber \\
\langle \varepsilon_t \rangle = \langle \nu_t \rangle = 0 \nonumber \\
\langle \varepsilon^2_t \rangle = \langle \nu^2_t \rangle = 1 \nonumber \\
\langle \varepsilon_t\nu_t \rangle = \rho_{\varepsilon\nu}
\label{SRCC}
\end{gather}
and the long-term memory processes are covered by ARFIMA(0,$d$,0) processes $\{x_t\}$ and $\{y_t\}$ with correlated error terms:
\begin{gather}
x_t=\sum_{n=0}^{+\infty}{a_n(d_1)\varepsilon_{t-n}} \nonumber \\
y_t=\sum_{n=0}^{+\infty}{a_n(d_2)\nu_{t-n}} \nonumber \\
a_n(d_i)=\frac{\Gamma(n+d_i)}{\Gamma(n+1)\Gamma(d_i)} \nonumber \\
\langle \varepsilon_t \rangle = \langle \nu_t \rangle = 0 \nonumber \\
\langle \varepsilon^2_t \rangle = \langle \nu^2_t \rangle = 1 \nonumber \\
\langle \varepsilon_t\nu_t \rangle = \rho_{\varepsilon\nu}
\label{LRCC}
\end{gather}

To discuss the basic properties of the test\footnote{R-project codes for the rescaled covariance test are available at \url{http://staff.utia.cas.cz/kristoufek/Ladislav_Kristoufek/Codes.html} or upon request from the author.}, we set $\theta_1=\theta_2=\theta$ and $d_1=d_2=d$ and we fix $q=30$. Note that the fractional differencing parameter $d$ is connected to the long-term memory Hurst exponent as $H=d+0.5$. For the short-range cross-correlated processes, we observe that the mean value is remarkably stable for parameters up to $\theta=0.7$ regardless of the correlation between error terms. For higher values, the statistic deviates which can be, however, attributed to the fact that we applied $q=30$ for estimation of the test statistic and that is evidently insufficient for such a strong memory. Interestingly, the mean value of the test statistic for $0\le \theta \le 0.7$ practically overlays with the testing statistic of the rescaled variance test \cite{Giraitis2003} which is defined as
\begin{equation}
U=\int_0^1\left(W^0_t\right)^2dt-\left(\int_0^1W^0_tdt\right)^2
\end{equation}
where $W^0_t$ is the standard Brownian bridge. Mean value of the statistic $U$ is equal to 1/12, which is represented by a red line in Fig. \ref{fig:RCT_1}. In the figure, we also show behavior of the standard deviation of the statistic. Even though it is evidently dependent on the correlation between error terms of the AR(1) processes, it is remarkably stable across different levels of $\theta$. Importantly, the variance decreases with increasing correlation between error terms which is a very desirable property. For the perfectly correlated error terms of the series, the standard deviation of the statistics even attains the levels for $U$ which is equal to $1/\sqrt{360}$. For the long-range cross-correlated processes, we observe that the mean value of the statistic increases with $d$ as expected. Again, the mean value is very stable with respect to the correlation of error terms. However, the variance of the estimator increases with $d$ parameter and is also dependent on the correlations between error terms. 

\section{Finite sample properties}

Even though the $M_{xy,T}(q)$ statistic shows some very desirable properties, we opt to base our decision in favor or against the alternative hypothesis based on the moving-block bootstrap (MBB) procedure \cite{Efron1979,Efron1993,Srinivas2000}, mainly due to dependence of the variance of the estimator on the correlations level. In the procedure, a bootstrapped series is obtained by separating the series into blocks of size $\zeta$ and shuffling the blocks, the parameter of interest is then estimated on the bootstrapped series for which the short-range dependence and the distributional properties of the original series are preserved. Based on $B$ bootstrapped estimates, the empirical confidence intervals for a specific level $\alpha$ and an empirical $p$-value are obtained. In the case of the rescaled covariance test, we work with a two-sided test with the null hypothesis of short-range cross-correlated processes against the alternative hypothesis of cross-persistence. 

To examine the size and power of the test, we use the same setup as in the previous section (Eqs. \ref{SRCC}-\ref{LRCC}). Specifically, we are interested in the finite sample properties of the rescaled covariance test for correlated, short-term correlated and long-term correlated processes with moderately and strongly correlated error terms. For the first case, we simply use a bivariate Gaussian noise series. For the second one, we utilize AR(1) processes with three levels of memory -- $\theta=0.1,0.5,0.8$ -- to control for weak, medium and strong cross-correlations. For the last one, we employ ARFIMA(0,$d$,0) processes with two levels of memory -- $d=0.1,0.4$ -- to discuss weak and strong power-law cross-correlations. For all previous cases, we discuss two levels of correlation between the error terms -- 0.5 and 0.9.

For correlated but not cross-correlated processes (Table \ref{tab:M}), we observe that the test is more precise with increasing correlation $\rho_{\varepsilon\nu}$ between error terms of the processes. For $\rho_{\varepsilon\nu}=0.9$, the size of the test practically matches the set significance levels. The size of the test gets better with increasing $q$ and practically does not vary with time series length $T$. Practically the same results are observed for the short-range cross-correlated processes as shown in Table \ref{tab:M_AR1}. The sizes practically overlay with the theoretical values of the significance levels. These are very strong results in favor of the rescaled covariance test as it is practically intact by even very strong short-term memory. The combination of the moving-block bootstrap and HAC-estimator of covariance is evidently able to sufficiently control for possible short-term memory biases in case of the RCT test.

For long-range cross-correlated processes, we compare cases when $H_x=H_y=0.6$ and $H_x=H_y=0.9$ to distinguish between weak and strong cross-persistence. We assume these values of $H_x$ and $H_y$ in the testing procedure. The power of the test is relatively low for the weak cross-persistence case (Table \ref{tab:M_ARFIMA1}). We, however, observe several interesting points. First, the power of the test is very similar regardless the correlation level between error terms. Second, the power of the test increases with the time series length. Third, the power increases rapidly with increasing $\alpha$. And fourthly, the power of the test even increases with an increasing $q$, which is caused by the $q^{\widehat{H_x}+\widehat{H_y}-1}$ factor in the testing statistic which well compensates for high $q$. For the strong cross-persistence (Table \ref{tab:M_ARFIMA2}), the power of the test increases considerably and the four features of the test are the same as in the previous case. As expected, the test is more powerful with increasing $\rho_{\varepsilon\nu}$, i.e. the cross-persistence is more stable. The power of the test increases to as high as 0.967 for some cases. The test thus shows very good statistical characteristics and is well able to distinguish between short-range and long-range cross-correlations.

\section{Application}

In financial economics, volatility is one of the most important variables as it is utilized in option pricing, portfolio analysis and risk management. In econophysics, volatility has been frequently studied due to its power-law nature (long-term memory, extreme events and aftershocks dynamics to name the most important ones). Studying the power-law cross-correlations in financial series thus naturally leads to the financial series connected to volatility. To utilize the proposed rescaled covariance test, we analyze two pairs of series which are of the main interest in finance -- volatility/returns and volatility/volume. Both pairs are interesting from the economics point of view -- volatility/return relationship is known as the leverage effect as negative returns are believed to be followed by increasing volatility \cite{Cont2001,Bollerslev2006}, and volatility/volume pair is interesting due to the fact that both variables are influenced by similar effects and one may influence the other \cite{Karpoff1987}. 

\begin{figure}[!htbp]
\begin{center}
\begin{tabular}{cc}
\includegraphics[width=70mm]{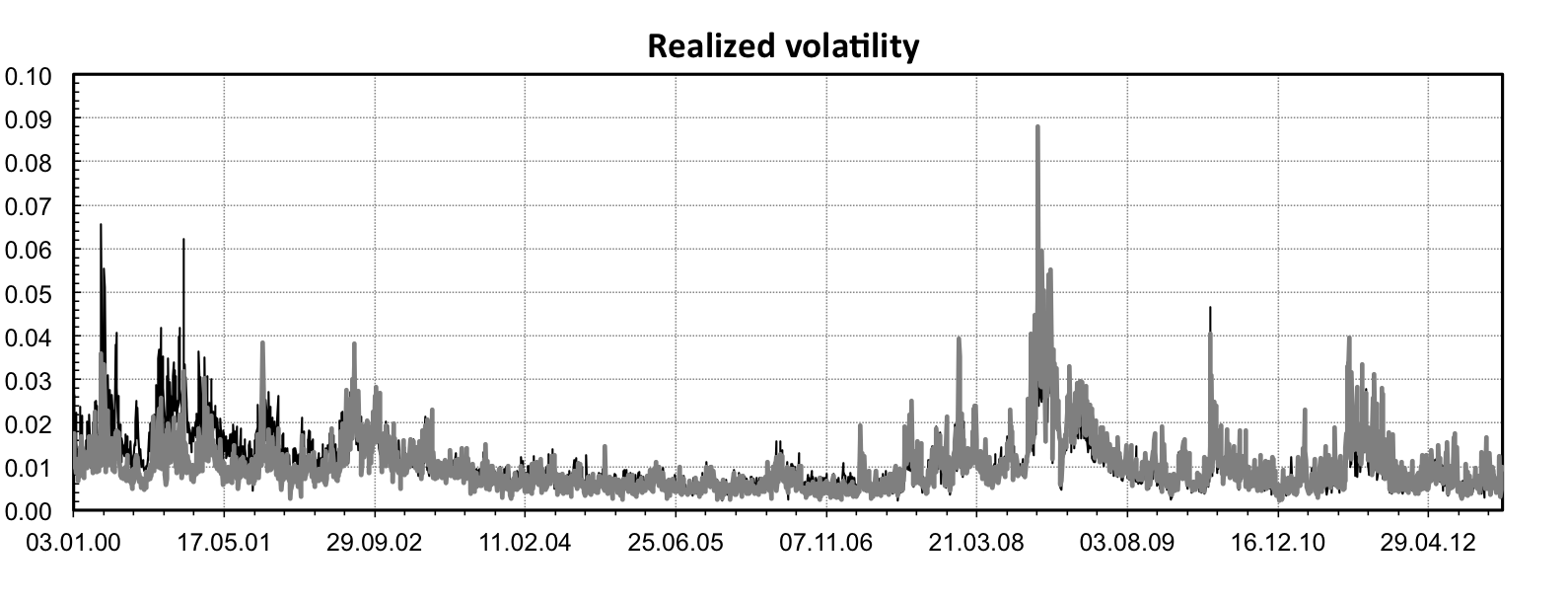}&\includegraphics[width=70mm]{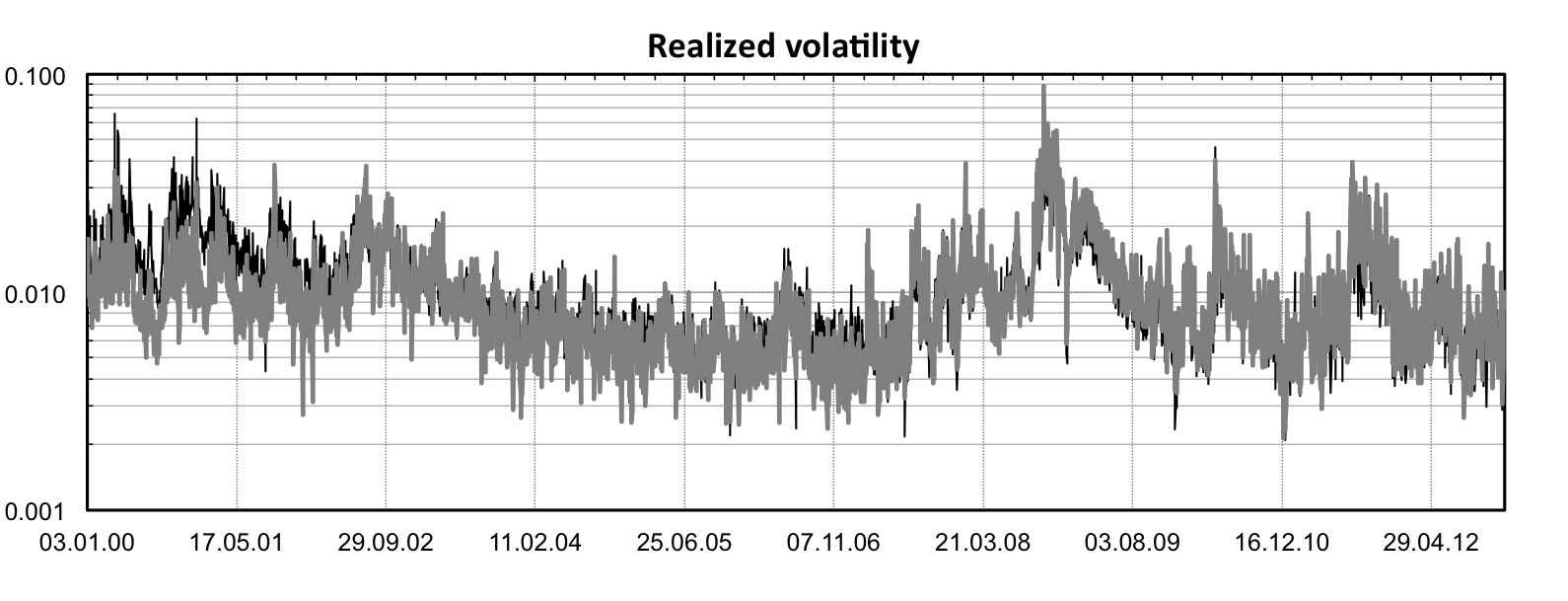}\\
\includegraphics[width=70mm]{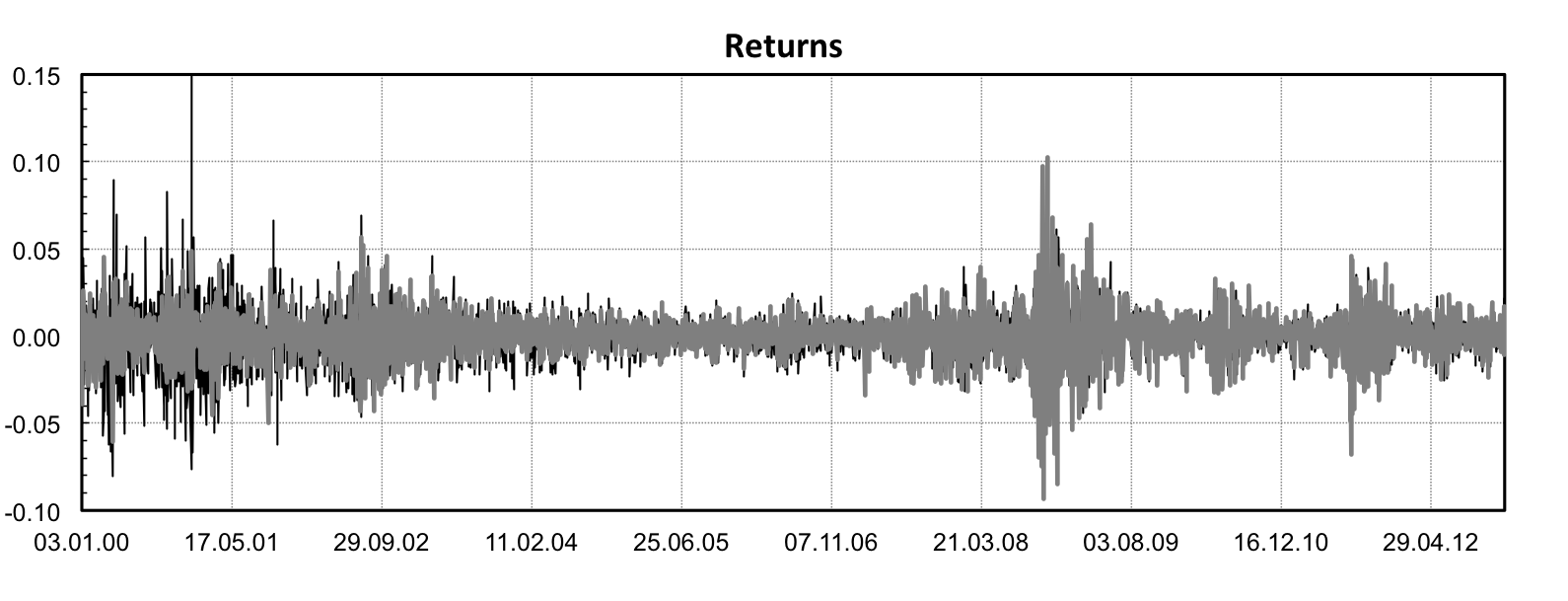}&\includegraphics[width=70mm]{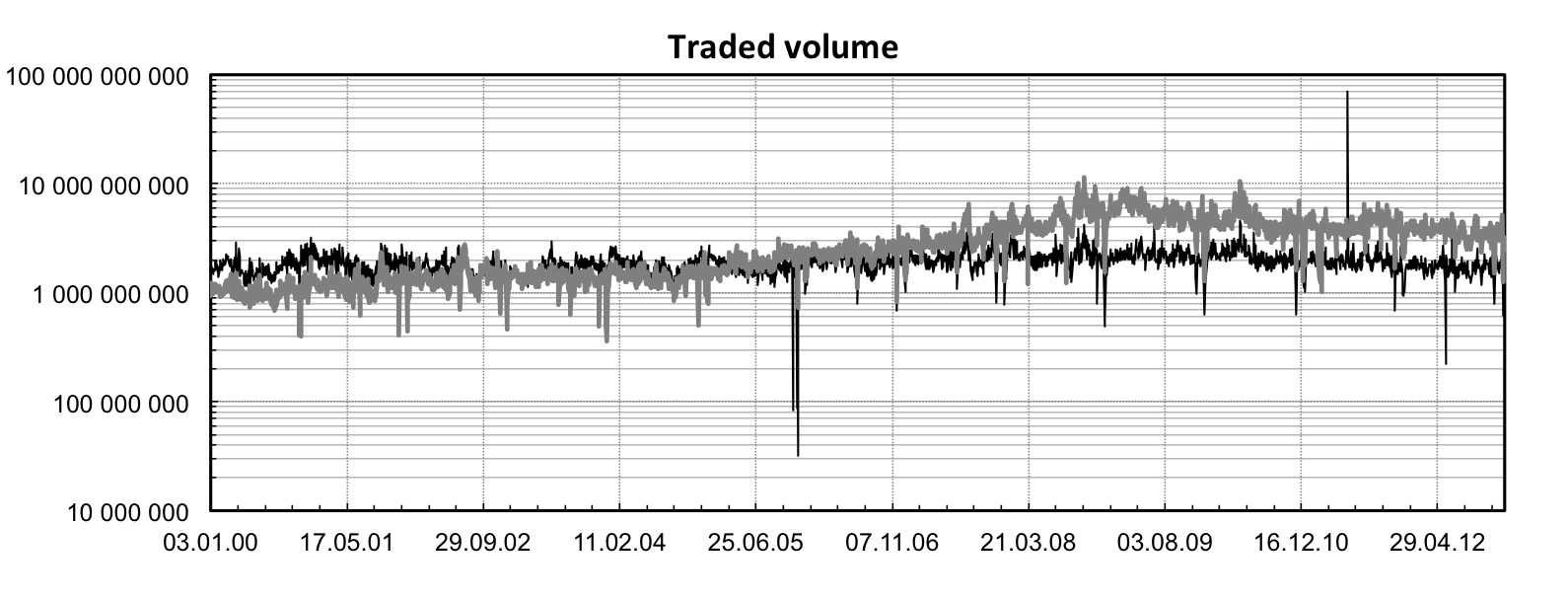}\\
\end{tabular}
\caption{\textbf{Volatility, returns and traded volume of NASDAQ-100 and S\&P500. }\footnotesize{Realized volatility (\textit{top left}), logarithmic realized volatility (\textit{top right}), logarithmic returns (\textit{bottom left}) and logarithmic traded volume (\textit{bottom right}) are shown for NASDAQ-100 (in black) and S\&P500 (in grey)}.\label{fig:TS1}}
\end{center}
\end{figure}

The volatility process is estimated with a use of the realized variance (volatility) approach, which employs the high-frequency data and yields consistent and efficient estimates of the true variance process \cite{Barndorff-Nielsen2002,Barndorff-Nielsen2002a,Hansen2006}. The realized variance is practically the uncentered second moment of the high-frequency series during a specific day. In our case, we use the 5 minutes frequency, which provides a good balance between efficiency and market microstructure noise bias. The realized variance is then defined as
\begin{equation}
\widehat{\sigma^2_{t,RV}}=\sum_{i=1}^{n}{r^2_{t,i}}
\end{equation}
where $r_{t,i}$ is a return of the $i$-th 5-minute interval during day $t$ and $n$ is the number of these 5-minute intervals for a given day. To overcome potential problems with non-standard distribution and non-negativity of the volatility series, we focus on the logarithmic volatility, i.e. the logarithm of the square root of the realized variance, which is standardly done in the literature \cite{Muzy2000}. In our analysis, we focus on two US indices -- NASDAQ-100 and S\&P500 -- between 1.1.2000 and 31.12.2012 (3245 and 3240 observations, respectively). In Fig. \ref{fig:TS1}, we observe that returns and volatility series for both indices practically overlap and the indices experienced very similar periods of increased volatility after the DotCom bubble of 2000 and an outburst of the Global Financial Crisis in 2007/2008. Development of the traded volume differs for the indices as the volume of the NASDAQ index has been quite stable during the analyzed period while the S\&P500 underwent an increasing exponential trend until the break of 2008 and 2009, stabilizing afterwards. To control for this development of the trading volume, we focus our analysis on the detrended logarithmic volume series.

\begin{figure}[!htbp]
\begin{center}
\begin{tabular}{cc}
\includegraphics[width=70mm]{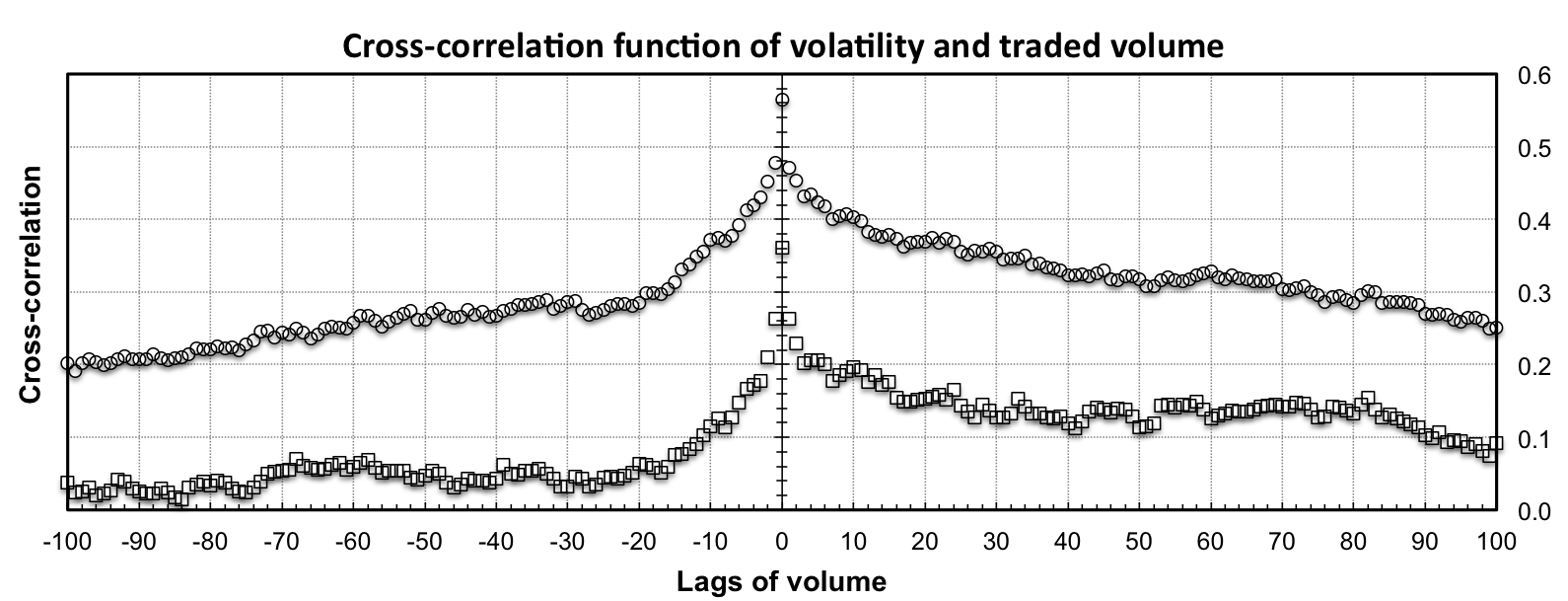}&\includegraphics[width=70mm]{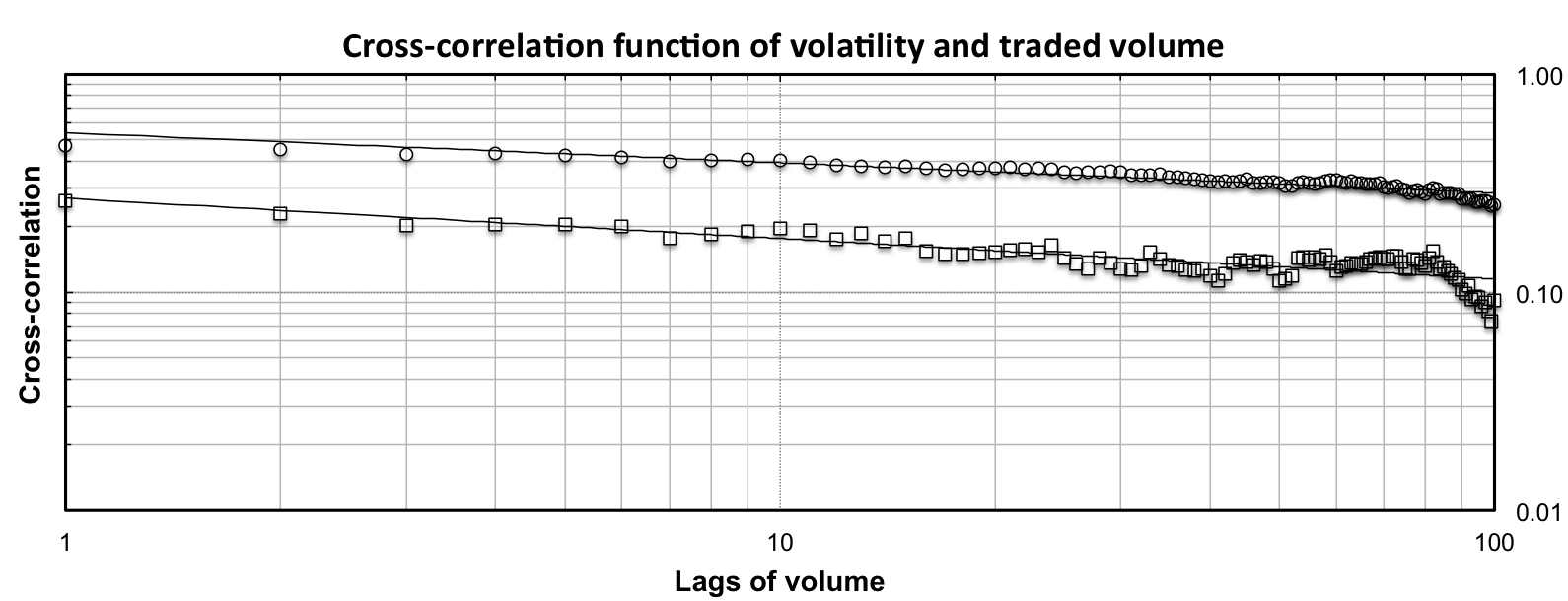}\\
\includegraphics[width=70mm]{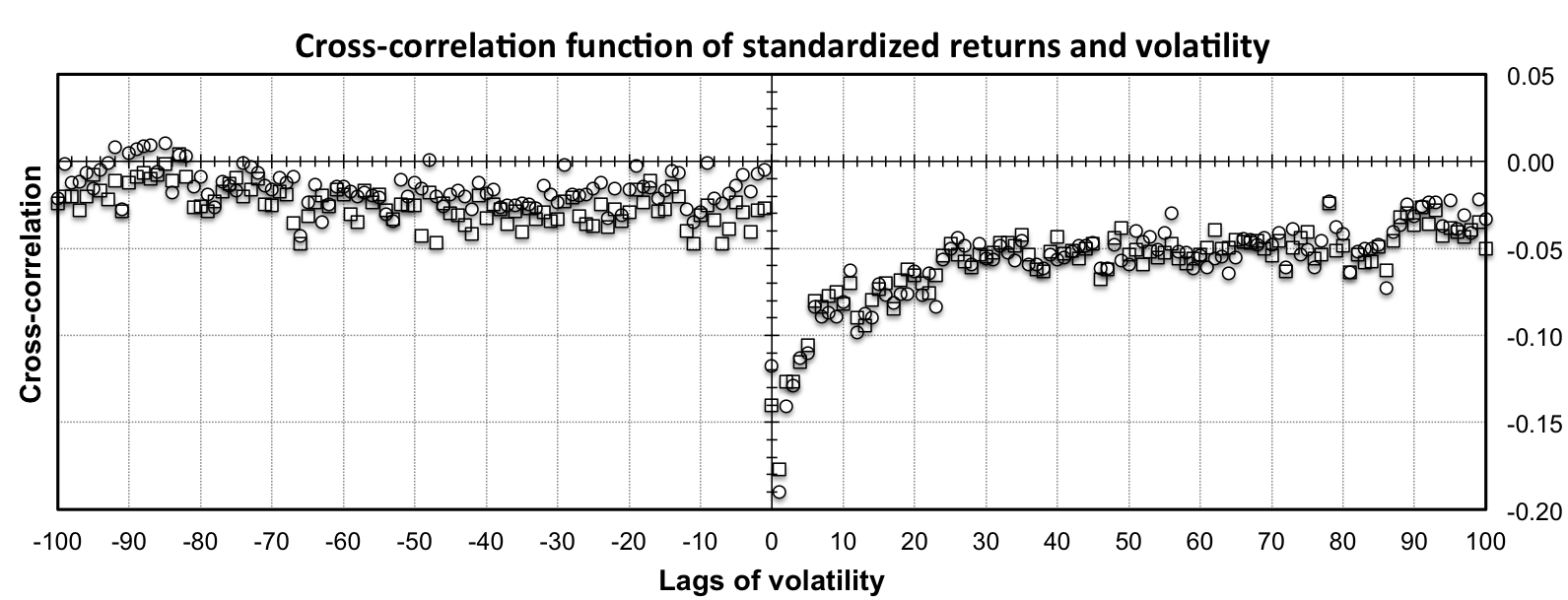}&\includegraphics[width=70mm]{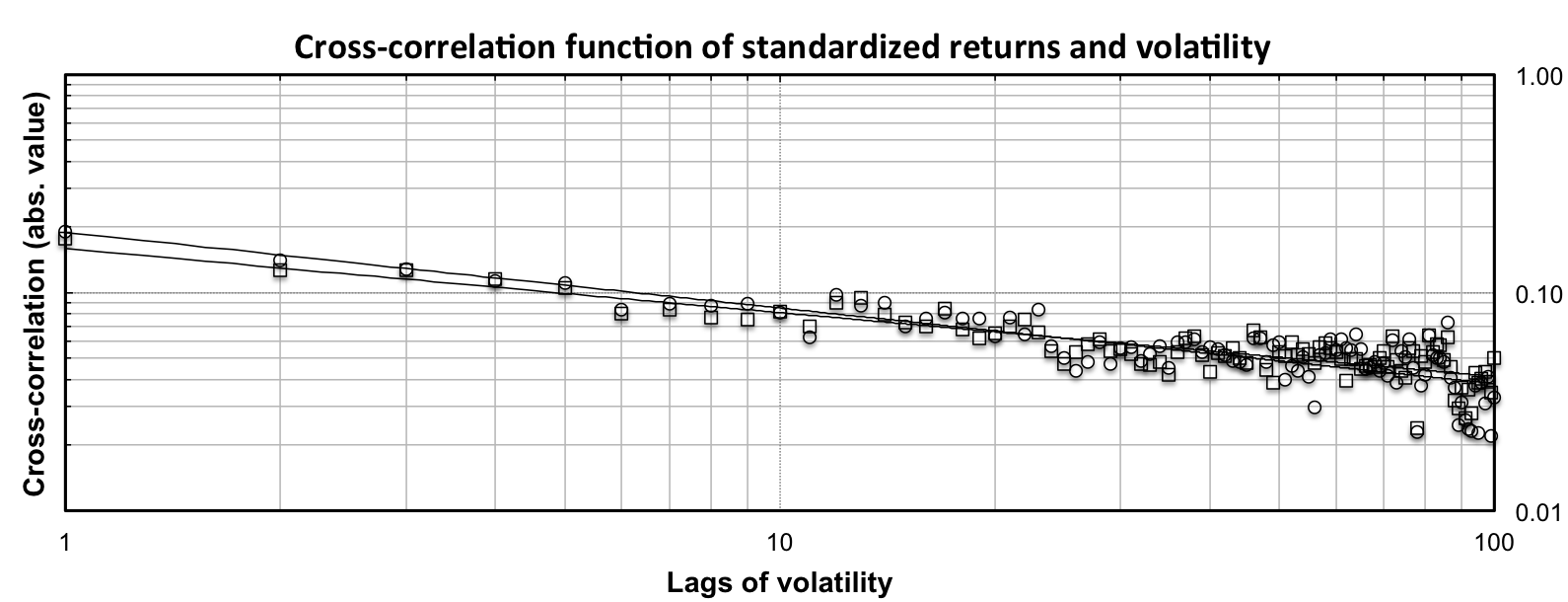}\\
\end{tabular}
\caption{\textbf{Cross-correlation functions for returns, volatility and traded volume of NASDAQ-100 and S\&P500. }\footnotesize{Cross-correlatios among volatility and traded volume (\textit{top left} and in log-log scale in \textit{top right}), and among returns and volatility (\textit{bottom left} and in log-log scale in \textit{bottom right}) are shown for NASDAQ-100 ($\square$) and S\&P500 ($\circ$)}.\label{fig:TS2}}
\end{center}
\end{figure}

Prior to turning to the results of the rescaled covariance test, we present the cross-correlation functions for both analyzed pairs in Fig. \ref{fig:TS2}. We observe that the relationships are very different from one another. Starting with the volatility/volume pair, we can see that positive cross-correlations are present for both halves of the cross-correlation function for both analyzed indices. For both, we find that the effect works in both directions. However, the effect of volatility on traded volume is more long-lasting than the other way around. Interestingly, the shape of the cross-correlation function is very similar for both indices but the level of correlations is approximately halved for NASDAQ-100 compared to the S\&P500 index. Nonetheless, a simple visual detection uncovers that the pair is a good candidate for the presence of LRCC. Such statement is further supported by visible power-law scaling of the right part of the cross-correlation function shown in the right panel of Fig. \ref{fig:TS2}. Turning to the returns/volatility pair, we can see a very different shape of the cross-correlation function which is strongly asymmetric. We observe a one-way effect from returns to volatility and not the other way around. Since the sample cross-correlations for the positive lags are all negative, it implies that positive (negative) returns cause, on statistical basis, decrease (increase) of volatility. This result is well in hand with the standard notion of the leverage effect in finance. Again, the decay of cross-correlations for positive lags is very slow and the pair is again a good candidate for the LRCC analysis which is visually supported by the power-law decay of the right part of the cross-correlation function illustrated in the right panel of Fig. \ref{fig:TS2}. We thus have two pairs suspected to be LRCC while one being positively and the other negatively cross-persistent.

\begin{figure}[!htbp]
\begin{center}
\begin{tabular}{cc}
\includegraphics[width=70mm]{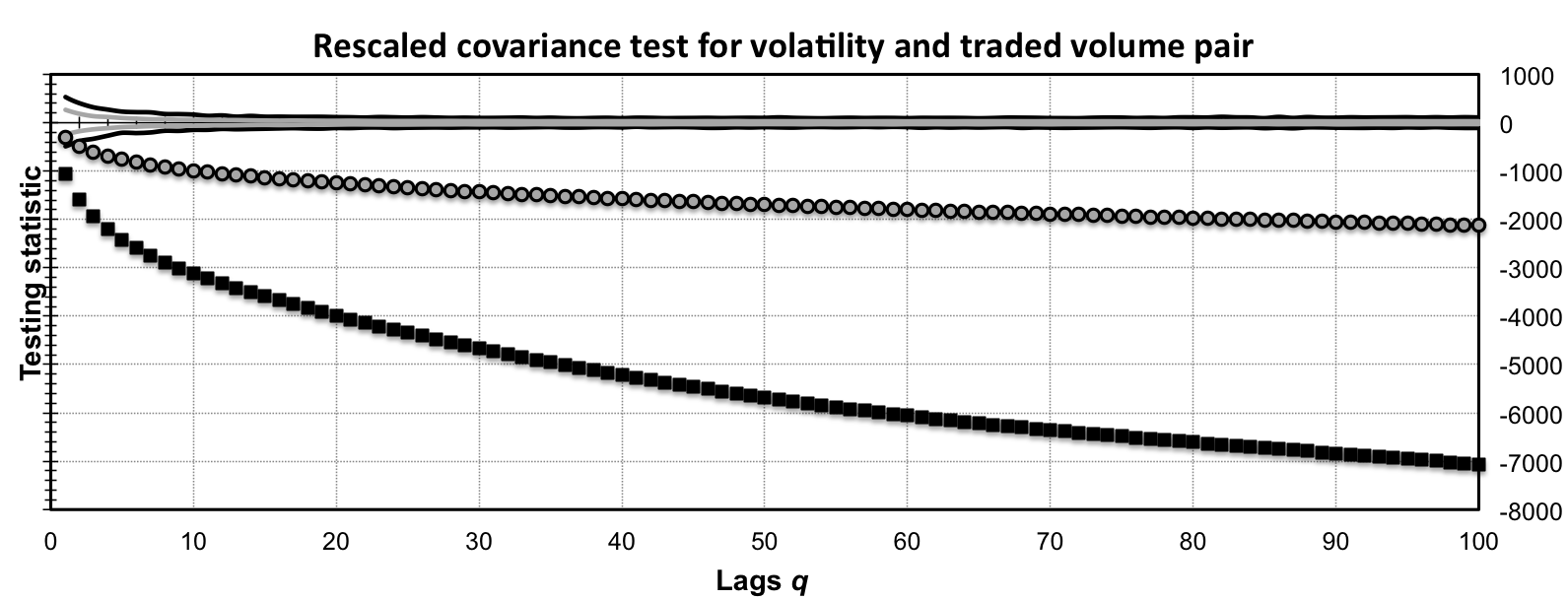}&\includegraphics[width=70mm]{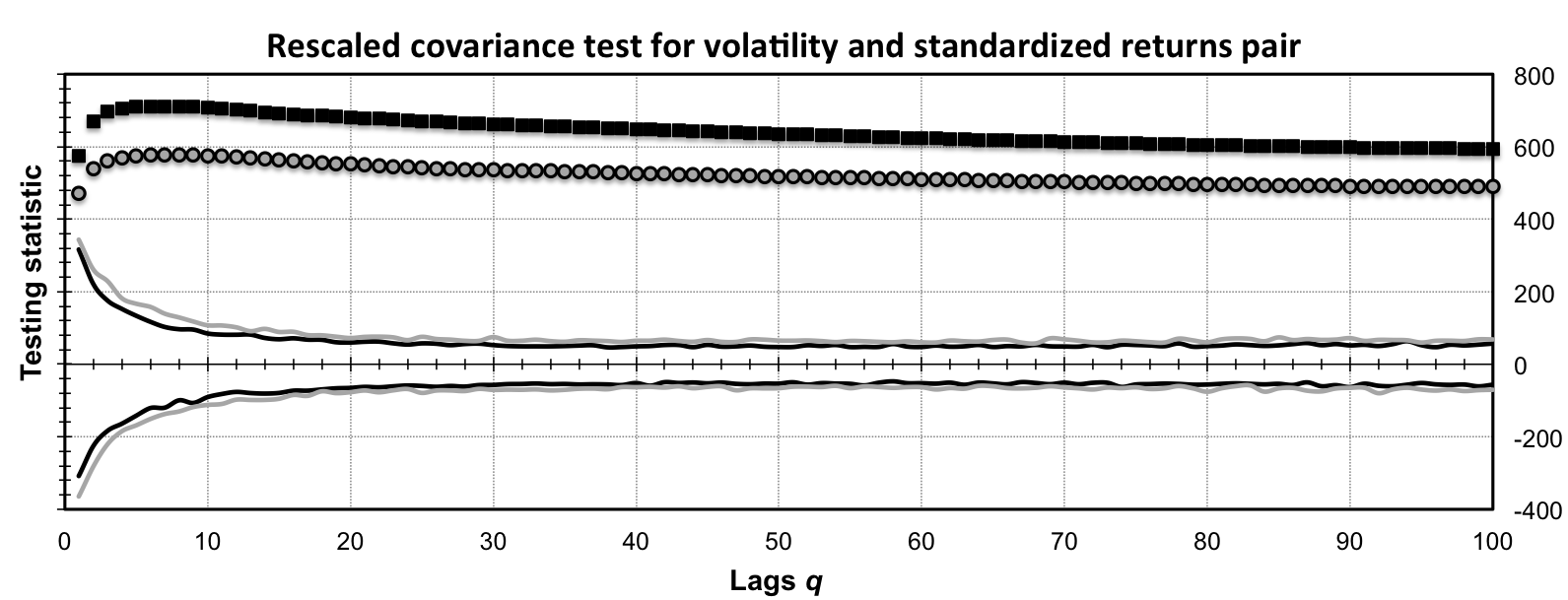}\\
\end{tabular}
\caption{\textbf{Rescaled covariance statistics $M_{xy,T}(q)$ for NASDAQ-100 and S\&P500. }\footnotesize{Testing statistics are shown for varying $q$ parameter between 1 and 100 to control for short-term memory. The statistics are shown for NASDAQ-100 ($\square$) and S\&P500 ($\circ$) and the 95\% confidence intervals are shown in solid lines (black for NASDAQ-100 and grey for S\&P500). If the testing statistics lay outside of the confidence intervals, the null hypothesis of no LRCC is rejected. The results are shown for the volatility-volume (\textit{left}) and returns-volatility (\textit{right}) pairs.}\label{fig:TS3}}
\end{center}
\end{figure}

Results of the rescaled covariance test for both pairs are summarized in Fig. \ref{fig:TS3}. In the figure, we present the testing statistic $M_{xy,T}(q)$ for parameter $q$ varying between 1 and 100 to see its behavior for different memory strengths. For the volatility/volume pair, we observe that the testing statistic is well below the critical values indicating statistically significant cross-persistence. This is true both for NASDAQ-100 and for S\&P500. The results are robust across different lags $q$ taken into consideration and evidently, the LRCC is not spuriously found due to the short-term memory bias. For the returns/volatility pair, we again find that there is a statistical evidence of long-range cross-correlations among returns and volatility. This is again true regardless the number of lags $q$ taken into consideration\footnote{We observe that the signs of the testing statistic are different for returns/volatility (positive) and volume/volatility (negative) pairs. For the former pair, this is caused by the fact that both the covariance of the partial sums and the covariance between original series are negative. And for the latter, the negativity indicates that even though both the volume and the volatility series are persistent, their partial sums follow local trends of opposite directions quite frequently. This stresses the need of the test to be two-sided.}. Both pairs are thus power-law cross-correlated according to the rescaled covariance test. 

\section{Conclusions}

We introduced a new test for detection of power-law cross-correlations among a pair of time series -- the rescaled covariance test. The test is based on a power-law divergence of the covariance of the partial sums of the LRCC processes. Together with a heteroskedasticity and auto-correlation robust (HAC) estimator of the long-term covariance, we developed a test with desirable statistical properties. As the application, we showed that the relationship between volatility and traded volume, and volatility and returns in the financial markets can be labeled as the one with power-law cross-correlations. Such test should be used as a starting point in the analysis of long-range cross-correlations prior to an estimation of bivariate long-term memory parameters.

\section*{Acknowledgements}

The support from the Grant Agency of Charles University (GAUK) under project $1110213$, Grant Agency of the Czech Republic (GACR) under projects P402/11/0948 and 402/09/0965, and project SVV 267 504 are gratefully acknowledged.

\bibliographystyle{plain}

\onecolumn

\begin{table}[htbp]
\centering
\caption[Size of $M_{xy,T}(q)$ statistic I]{\textbf{Size of $M_{xy,T}(q)$ statistic I.} Monte-Carlo-based test size for 1,000 replications of processes $x_t=\varepsilon_t$ and $y_t=\nu_t$ with different correlations $\rho_{\varepsilon\nu}$.\label{tab:M}}
\begin{tabular}{c|c|ccc|ccc}
&&&$\rho=0.5$&&&$\rho=0.9$&\\
\hline
&&$\alpha=0.01$&$\alpha=0.05$&$\alpha=0.1$&$\alpha=0.01$&$\alpha=0.05$&$\alpha=0.1$\\
\hline \hline
&$q=1$&0.011&0.045&0.092&0.011&0.050&0.099\\
$T=500$&$q=5$&0.009&0.042&0.092&0.010&0.050&0.099\\
&$q=10$&0.011&0.042&0.090&0.011&0.052&0.102\\
&$q=30$&0.011&0.042&0.090&0.011&0.052&0.102\\
\hline
&$q=1$&0.011&0.048&0.101&0.014&0.062&0.094\\
$T=1000$&$q=5$&0.012&0.052&0.101&0.014&0.060&0.094\\
&$q=10$&0.011&0.053&0.100&0.014&0.053&0.095\\
&$q=30$&0.011&0.053&0.100&0.014&0.053&0.095\\
\hline
&$q=1$&0.014&0.047&0.100&0.012&0.049&0.101\\
$T=5000$&$q=5$&0.014&0.048&0.102&0.012&0.050&0.100\\
&$q=10$&0.014&0.048&0.098&0.012&0.050&0.099\\
&$q=30$&0.014&0.048&0.098&0.012&0.050&0.099\\
\end{tabular}
\end{table}

\begin{table}[htbp]
\centering
\caption[Size of $M_{xy,T}(q)$ statistic II]{\textbf{Size of $M_{xy,T}(q)$ statistic II.} Monte-Carlo-based test size for 1,000 replications of two AR(1) processes with $\theta_x=\theta_y=0.1$ and different correlations $\rho_{\varepsilon\nu}$.\label{tab:M_AR1}}
\begin{tabular}{c|c|ccc|ccc}
&&&$\rho=0.5$&&&$\rho=0.9$&\\
\hline
&&$\alpha=0.01$&$\alpha=0.05$&$\alpha=0.1$&$\alpha=0.01$&$\alpha=0.05$&$\alpha=0.1$\\
\hline \hline
&$q=1$&0.006&0.045&0.109&0.009&0.036&0.084\\
$T=500$&$q=5$&0.005&0.048&0.104&0.009&0.038&0.082\\
&$q=10$&0.006&0.048&0.108&0.007&0.034&0.085\\
&$q=30$&0.006&0.048&0.108&0.007&0.034&0.085\\
\hline
&$q=1$&0.013&0.061&0.102&0.018&0.049&0.093\\
$T=1000$&$q=5$&0.010&0.063&0.104&0.017&0.049&0.087\\
&$q=10$&0.010&0.058&0.105&0.018&0.048&0.090\\
&$q=30$&0.010&0.058&0.105&0.018&0.048&0.090\\
\hline
&$q=1$&0.014&0.054&0.117&0.011&0.050&0.109\\
$T=5000$&$q=5$&0.014&0.053&0.114&0.012&0.050&0.110\\
&$q=10$&0.014&0.051&0.115&0.012&0.052&0.109\\
&$q=30$&0.014&0.051&0.115&0.012&0.052&0.109\\
\end{tabular}
\end{table}

\begin{table}[htbp]
\centering
\caption[Size of $M_{xy,T}(q)$ statistic III]{\textbf{Size of $M_{xy,T}(q)$ statistic III.} Monte-Carlo-based test size for 1,000 replications of two AR(1) processes with $\theta_x=\theta_y=0.5$ and different correlations $\rho_{\varepsilon\nu}$.\label{tab:M_AR2}}
\begin{tabular}{c|c|ccc|ccc}
&&&$\rho=0.5$&&&$\rho=0.9$&\\
\hline
&&$\alpha=0.01$&$\alpha=0.05$&$\alpha=0.1$&$\alpha=0.01$&$\alpha=0.05$&$\alpha=0.1$\\
\hline \hline
&$q=1$&0.006&0.044&0.101&0.012&0.046&0.095\\
$T=500$&$q=5$&0.003&0.043&0.095&0.012&0.047&0.084\\
&$q=10$&0.005&0.044&0.092&0.009&0.046&0.083\\
&$q=30$&0.005&0.044&0.092&0.009&0.046&0.083\\
\hline
&$q=1$&0.011&0.057&0.103&0.012&0.049&0.104\\
$T=1000$&$q=5$&0.009&0.053&0.099&0.012&0.046&0.096\\
&$q=10$&0.008&0.052&0.093&0.012&0.043&0.096\\
&$q=30$&0.008&0.052&0.093&0.012&0.043&0.096\\
\hline
&$q=1$&0.006&0.047&0.090&0.015&0.053&0.106\\
$T=5000$&$q=5$&0.006&0.042&0.083&0.013&0.055&0.107\\
&$q=10$&0.005&0.043&0.079&0.012&0.056&0.106\\
&$q=30$&0.005&0.043&0.079&0.012&0.056&0.106\\
\end{tabular}
\end{table}

\begin{table}[htbp]
\centering
\caption[Size of $M_{xy,T}(q)$ statistic IV]{\textbf{Size of $M_{xy,T}(q)$ statistic IV.} Monte-Carlo-based test size for 1,000 replications of two AR(1) processes with $\theta_x=\theta_y=0.8$ and different correlations $\rho_{\varepsilon\nu}$.\label{tab:M_AR3}}
\begin{tabular}{c|c|ccc|ccc}
&&&$\rho=0.5$&&&$\rho=0.9$&\\
\hline
&&$\alpha=0.01$&$\alpha=0.05$&$\alpha=0.1$&$\alpha=0.01$&$\alpha=0.05$&$\alpha=0.1$\\
\hline \hline
&$q=1$&0.019&0.075&0.135&0.010&0.048&0.104\\
$T=500$&$q=5$&0.013&0.063&0.120&0.008&0.048&0.100\\
&$q=10$&0.011&0.058&0.116&0.009&0.047&0.094\\
&$q=30$&0.011&0.058&0.116&0.009&0.047&0.094\\
\hline
&$q=1$&0.020&0.068&0.130&0.014&0.050&0.097\\
$T=1000$&$q=5$&0.015&0.059&0.121&0.012&0.045&0.085\\
&$q=10$&0.012&0.054&0.110&0.011&0.047&0.083\\
&$q=30$&0.012&0.054&0.110&0.011&0.047&0.083\\
\hline
&$q=1$&0.017&0.072&0.120&0.022&0.065&0.108\\
$T=5000$&$q=5$&0.016&0.064&0.111&0.017&0.054&0.104\\
&$q=10$&0.013&0.058&0.104&0.017&0.053&0.102\\
&$q=30$&0.013&0.058&0.104&0.017&0.053&0.102\\
\end{tabular}
\end{table}

\begin{table}[htbp]
\centering
\caption[Power of $M_{xy,T}(q)$ statistic I]{\textbf{Power of $M_{xy,T}(q)$ statistic I.} Monte-Carlo-based test power for 1,000 replications of two ARFIMA(0,$d$,0) processes with $d_x=d_y=0.1$ and different correlations $\rho_{\varepsilon\nu}$.\label{tab:M_ARFIMA1}}
\begin{tabular}{c|c|ccc|ccc}
&&&$\rho=0.5$&&&$\rho=0.9$&\\
\hline
&&$\alpha=0.01$&$\alpha=0.05$&$\alpha=0.1$&$\alpha=0.01$&$\alpha=0.05$&$\alpha=0.1$\\
\hline \hline
&$q=1$&0.018&0.087&0.148&0.029&0.094&0.141\\
$T=500$&$q=5$&0.081&0.184&0.275&0.103&0.196&0.278\\
&$q=10$&0.117&0.232&0.343&0.142&0.254&0.344\\
&$q=30$&0.117&0.232&0.343&0.142&0.254&0.344\\
\hline
&$q=1$&0.030&0.111&0.172&0.023&0.090&0.166\\
$T=1000$&$q=5$&0.097&0.205&0.295&0.094&0.215&0.312\\
&$q=10$&0.135&0.252&0.349&0.155&0.283&0.369\\
&$q=30$&0.135&0.252&0.349&0.155&0.283&0.369\\
\hline
&$q=1$&0.091&0.200&0.283&0.090&0.201&0.282\\
$T=5000$&$q=5$&0.187&0.320&0.409&0.195&0.342&0.438\\
&$q=10$&0.233&0.368&0.466&0.235&0.399&0.500\\
&$q=30$&0.233&0.368&0.466&0.235&0.399&0.500\\
\end{tabular}
\end{table}

\begin{table}[htbp]
\centering
\caption[Power of $M_{xy,T}(q)$ statistic II]{\textbf{Power of $M_{xy,T}(q)$ statistic II.} Monte-Carlo-based test power for 1,000 replications of two ARFIMA(0,$d$,0) processes with $d_x=d_y=0.4$ and different correlations $\rho_{\varepsilon\nu}$.\label{tab:M_ARFIMA2}}
\begin{tabular}{c|c|ccc|ccc}
&&&$\rho=0.5$&&&$\rho=0.9$&\\
\hline
&&$\alpha=0.01$&$\alpha=0.05$&$\alpha=0.1$&$\alpha=0.01$&$\alpha=0.05$&$\alpha=0.1$\\
\hline \hline
&$q=1$&0.111&0.229&0.318&0.147&0.272&0.356\\
$T=500$&$q=5$&0.649&0.725&0.768&0.734&0.797&0.839\\
&$q=10$&0.772&0.830&0.862&0.869&0.904&0.924\\
&$q=30$&0.772&0.830&0.862&0.869&0.904&0.924\\
\hline
&$q=1$&0.205&0.339&0.421&0.255&0.371&0.464\\
$T=1000$&$q=5$&0.697&0.774&0.814&0.747&0.813&0.846\\
&$q=10$&0.817&0.867&0.891&0.857&0.893&0.914\\
&$q=30$&0.817&0.867&0.891&0.857&0.893&0.914\\
\hline
&$q=1$&0.464&0.584&0.636&0.584&0.685&0.737\\
$T=5000$&$q=5$&0.823&0.878&0.899&0.892&0.922&0.934\\
&$q=10$&0.898&0.922&0.933&0.934&0.958&0.967\\
&$q=30$&0.898&0.922&0.933&0.934&0.958&0.967\\
\end{tabular}
\end{table}

\section*{Appendix}

\subsection*{Proof to ``Partial sum covariance scaling'' proposition}
Using the zero mean and stationarity properties of processes $\{x_t\}$ and $\{y_t\}$, we can write the covariance of the partial sums as 
\begin{multline}
\text{Cov}(X_n,Y_n)=\langle X_nY_n \rangle = \sigma_x\sigma_y \Big(n\rho_{xy}(0)+\sum_{k=1}^{n-1}(n-k)(\rho_{xy}(k)+\rho_{xy}(-k))\Big) \\
\propto n\rho_{xy}(0)+\sum_{k=1}^{n-1}(n-k)(\rho_{xy}(k)+\rho_{xy}(-k)).
\end{multline}
Now, assuming that $\rho_{xy}(k)$ is symmetric for $k>0$ and $k<0$, we have
\begin{equation}
\text{Cov}(X_n,Y_n)\propto n\rho_{xy}(0)+ n\sum_{k=1}^{n-1}{\rho_{xy}(k)}- \sum_{k=1}^{n-1}{k\rho_{xy}(k)}.
\label{eq:sym}
\end{equation}
Using the LRCC definition and approximating the infinite sums with definite integrals according to the Euler--MacLaurin integration formula \cite{Euler1738,MacLaurin1742}, we get
\begin{equation}
\label{eq:int1}
n\sum_{k=1}^{n-1}{\rho_{xy}(k)}\propto n\sum_{k=1}^{n-1}k^{-\gamma_{xy}}\approx n \int_1^n{k^{-\gamma_{xy}}dk}\propto n^{2-\gamma_{xy}},
\end{equation}
\begin{equation}
\label{eq:int2}
\sum_{k=1}^{n-1}{k\rho_{xy}(k)}\propto \sum_{k=1}^{n-1}k^{1-\gamma_{xy}}\approx\int_1^n{k^{1-\gamma_{xy}}dk}\propto n^{2-\gamma_{xy}}.
\end{equation}
Finally, we use that the linear growth of $n\rho_{xy}(0)$ is asymptotically dominated by the power-law growth in the latter terms, i.e. using the l'H\^{o}pital's rule we have
\begin{equation}
\lim_{n\rightarrow +\infty}{\frac{n^{2-\gamma_{xy}}}{n\rho_{xy}(0)}}=\lim_{n\rightarrow +\infty}{\frac{(2-\gamma_{xy})n^{1-\gamma_{xy}}}{\rho_{xy}(0)}}=+\infty\text{ for }0<\gamma_{xy}<1
\end{equation}
and we get
\begin{equation}
\label{eq:crossHurst}
\text{Cov}(X_n,Y_n)\propto n^{2-\gamma_{xy}}\text{ as }n\rightarrow +\infty.
\end{equation}
Note that the substitutions in Eqs. \ref{eq:int1} and \ref{eq:int2} from $\sum_{k=1}^{n-1}{\rho_{xy}(k)}$ to $\sum_{k=1}^{n-1}{k^{-\gamma_{xy}}}$ are done for $k$ between $1$ and $n-1$ without a loss on generality as we are interested in the asymptotic properties of Cov($X_n,Y_n$).

Further, we have $2H_{xy}=2-\gamma_{xy}$ so that 
\begin{equation}
\label{eq:HurstGamma}
H_{xy}=1-\frac{\gamma_{xy}}{2}.
\end{equation}

For the asymmetric cross-correlation function, the results do not differ significantly. We have
\begin{equation}
\text{Cov}(X_n,Y_n)\approx n\rho_{xy}(0)+\underbrace{n\sum_{k=1}^{n-1}{k^{-\gamma^1_{xy}}}-\sum_{k=1}^{n-1}{k^{-\gamma^1_{xy}+1}}}_{\propto n^{2-\gamma^1_{xy}}}+\underbrace{n\sum_{k=1}^{n-1}{k^{-\gamma^2_{xy}}}-\sum_{k=1}^{n-1}{k^{-\gamma^2_{xy}+1}}}_{\propto n^{2-\gamma^2_{xy}}},
\end{equation}
where the approximate proportionality comes from Eqs. \ref{eq:int1} and \ref{eq:int2}. Asymptotically, the power-law scaling is dominated by the higher exponent, i.e. the lower $\gamma_{xy}$. For $\gamma_{xy}^1<\gamma_{xy}^2$, we have $\text{Cov}(X_n,Y_n)\sim n^{2-\gamma_{xy}^1}$ and vice versa. Note that the lower $\gamma_{xy}$ is connected to the higher bivariate Hurst exponent $H_{xy}$ which implies that the scaling of covariances is dominated by the stronger cross-persistence. $\square$\\

\subsection*{Proof to ``Diverging limit of covariance of partial sums'' proposition}
We have
\begin{multline}
\label{eq:PSlim1}
\lim_{n\rightarrow +\infty}{\frac{\text{Cov}(X_n,Y_n)}{n}}\propto \lim_{n\rightarrow +\infty}\frac{n^{2H_{xy}}}{n}=\lim_{n\rightarrow +\infty}\frac{n^{2-\gamma_{xy}}}{n}=\\
\lim_{n\rightarrow +\infty}n^{1-\gamma_{xy}}=+\infty\text{ for }0<\gamma_{xy}<1. \square
\end{multline}

\subsection*{Proof to ``Converging limit of covariance of partial sums'' proposition}
In accordance with the proof for the LRCC case, we assume a symmetric cross-correlation function\footnote{For an asymmetric case, the proof is parallel.} so that we can write
\begin{equation}
\text{Cov}(X_n,Y_n)\propto n\rho_{xy}(0)+ n\sum_{k=1}^{n-1}{\rho_{xy}(k)}- \sum_{k=1}^{n-1}{k\rho_{xy}(k)}.
\end{equation}
It holds that
\begin{equation}
\label{eq:PSlimit_a}
\lim_{n\rightarrow +\infty}{\frac{\text{Cov}(X_n,Y_n)}{n}} \propto \lim_{n\rightarrow +\infty}{\left(\rho_{xy}(0)+\sum_{k=1}^{n-1}{\rho_{xy}(k)}- \frac{1}{n}\sum_{k=1}^{n-1}{k\rho_{xy}(k)}\right)}.
\end{equation}
Solving the sums separately with a use of short-range cross-correlations definition, we get
\begin{equation}
\sum_{k=1}^{n-1}{\rho_{xy}(k)}\propto \sum_{k=1}^{n-1}{\exp\left(-\frac{k}{\delta}\right)} \propto \frac{1-\exp\left(-\frac{n}{\delta}\right)}{1-\exp\left(-\frac{1}{\delta}\right)}
\end{equation}
\begin{equation}
\sum_{k=1}^{n-1}{k\rho_{xy}(k)} \propto \sum_{k=1}^{n-1}{k\exp\left(-\frac{k}{\delta}\right)} = \exp\left(-\frac{1}{\delta}\right)-n\exp\left(-\frac{n}{\delta}\right)+(n-1)\exp\left(-\frac{n+1}{\delta}\right)
\end{equation}
Substituting back, we obtain
\begin{multline}
\lim_{n\rightarrow +\infty}{\frac{\text{Cov}(X_n,Y_n)}{n}} \propto \lim_{n\rightarrow +\infty}\Bigg[\rho_{xy}(0)+\frac{1-\exp\left(-\frac{n}{\delta}\right)}{1-\exp\left(-\frac{1}{\delta}\right)}-\frac{\exp\left(-\frac{1}{\delta}\right)}{n}+\\
\frac{n}{n}\exp\left(-\frac{n}{\delta}\right)+\frac{n-1}{n}\exp\left(-\frac{n+1}{\delta}\right)\Bigg]=\rho_{xy}(0)+\frac{1}{1-\exp\left(-\frac{1}{\delta}\right)}
\end{multline}
and the limit evidently converges for $0\le \delta < +\infty$ which concludes the proof. $\square$ \\

\end{document}